\documentclass[11pt]{article}

\usepackage{amsmath,amssymb}

\usepackage[margin=2.5cm]{geometry}

\usepackage{float}
\usepackage{graphicx,caption,subcaption}
\usepackage{color}
\graphicspath{{./Figures_jpeg_small/}}

\title{Security of two-state and four-state practical quantum bit-commitment
protocols}
\author{Ricardo Loura\textsuperscript{1,2}, Du\v{s}an Arsenovi\'{c}\textsuperscript{3}, Nikola Paunkovi\'{c}\textsuperscript{1,2}\thanks{npaunkov@math.tecnico.ulisboa.pt},\\ Du\v{s}ka B. Popovi\'{c}\textsuperscript{3} and Slobodan Prvanovi\'{c}\textsuperscript{3}}
\date{}

\begin{document}
\maketitle

\noindent\textsuperscript{1} Instituto de Telecomunica\c{c}\~{o}es, Av. Rovisco Pais 1049-001, Lisboa, Portugal\\
\textsuperscript{2} Departamento de Matem\' atica, Instituto Superior T\' ecnico, Universidade de Lisboa, Av. Rovisco Pais 1049-001, Lisboa, Portugal\\
\textsuperscript{3} Institute of Physics, University of Belgrade, Pregrevica 118, 11080 Belgrade, Serbia

\begin{abstract}
	We study cheating strategies against a practical four-state quantum bit-commitment protocol [A. Danan and L. Vaidman, Quant. Info. Proc. {\bf 11}, 769 (2012)] and its two-state variant [R. Loura {\em et al.}, Phys. Rev. A {\bf 89}, 052336 (2014)] when the underlying quantum channels are noisy and the cheating party is constrained to using single-qubit measurements only. We show that simply inferring the transmitted photons' states by using the Breidbart basis, optimal for ambiguous (minimum-error) state discrimination, does not directly produce an optimal cheating strategy for this bit-commitment protocol. We introduce a strategy, based on certain post-measurement processes, and show it to have better chances at cheating than the direct approach. We also study to what extent sending forged geographical coordinates helps a dishonest party in breaking the binding security requirement. Finally, we investigate the impact of imperfect single-photon sources in the protocols. Our study shows that, in terms of the resources used, the four-state protocol is advantageous over the two-state version. The analysis performed can be straightforwardly generalised to any finite-qubit measurement, with the same qualitative results.
\end{abstract}

\section{Introduction}

With the accelerating development of quantum information theory, new cryptographic protocols and primitives arise every day alongside new engineering techniques to turn them into reality. Naturally, due to the resemblance to classical cryptography and its implementations, new challenges and new difficulties appear once one moves from theory to practice. Quantifying the impact of these difficulties and studying their solutions is thus of utmost importance if quantum cryptography is to become a widespread reality. As a step towards this goal, we here study the feasibility of several different attacks on a practical two-state quantum bit-commitment (BC) protocol \cite{Loura14,alm:etal:14,alm:etal:15} and its original four-state variant \cite{Danan12} in a noisy environment. By practical, we mean that cheating is subjected to current technological constraints: the lack of long-term quantum memories and, in the case of the currently predominant optical realisations, the non-existence of photon non-demolition measurements. Thus, an agent (in our case Alice) is forced to perform measurements as soon as (s)he receives the qubits encoded in, say, photon polarisation. Various practical cryptographic protocols based on bounded and/or noisy memories have been studied~\cite{weh:sch:ter:08,sch:ter:weh:11,koe:weh:wul:12,ng:jos:min:kur:weh:12}, in order to avoid the Lo-Chau~\cite{lo:chau:97} and Mayers~\cite{may:97} no-go theorem regarding non-relativistic implementations of quantum bit-commitment schemes. (see also a recent proposal of a computationally secure oblivious transfer protocol~\cite{sou:mat:ada:pau:14}).

We start by reviewing the protocol's steps, including a careful definition of the verification procedure, in Section \ref{prtcl_def}. In Section \ref{opt_cheat} we discuss the so-called {\em cheating observable}, given by the Breidbart basis, and show that the na\"{i}ve approach, studied in \cite{Loura14}, can be perfected. In Section \ref{fake_dist}, we analyse the so-called {\em faking-distance attack}, based on providing wrong geographical information to the verifier. Finally, Section \ref{sing_phots} is dedicated to the so-called {\em beam-splitter attack}, which takes advantage of the fact that no perfect single-photon sources exist, much to the resemblance of BB84's photon-number splitting attack. We compare the enhanced attack with its counterpart in the four-state variant.

\section{The protocol}
\label{prtcl_def}

Commitment protocols, initially proposed in 1981 by Manuel Blum \cite{Blum81} and simultaneously by Shamir, Rivest and Adleman \cite{Shamir81}, are important current cryptographic primitives that serve as building blocks for several cryptographic and computational protocols, such as zero-knowledge proofs~\cite{gol:mic:wig:91} and multi-party computation~\cite{lin:07}. Simply put, a commitment scheme is a sequence of steps between a prover (Alice) and a verifier (Bob) that allows them to essentially roll a dice in a fair way over a communication channel. In the first step (the {\em commitment phase}), Alice makes a choice from a set of possible choices ($1,\dots ,n$ in the case of an actual $n$-sided dice). In the second step (the {\em revealing phase}), Bob asks Alice to reveal her choice and to \emph{prove} that it was indeed her initial choice. In practice, these two steps are not done immediately after one another, but are instead crucial parts of a larger protocol. Naturally, the scheme must guarantee that between the two phases Alice cannot change her mind (called the {\em binding} requirement), and that Bob cannot learn Alice's commitment choice before she reveals it (called the {\em concealing} requirement). For a more detailed description and examples of commitment schemes, see \cite{Loura14}. If Alice's choice is bound to be either $0$ or $1$, then we say that the commitment scheme is a bit-commitment scheme. Furthermore, if the protocol requires quantum communication of any kind, we say that it is a quantum commitment scheme.

In this work we are interested in a particular quantum bit commitment scheme initially presented in \cite{Loura14}, building on the earlier works of Danan and Vaidman in \cite{Danan12}. We begin with a brief review of the protocol's definition. A detailed version of what follows, along with a theoretical security analysis, can be found in \cite{Loura14} and in \cite{alm:etal:15}. Note, however, that in the present paper we use a slightly different notation. In fact, the works \cite{Loura14} and \cite{alm:etal:15} focus exclusively on the two-state protocol, whereas we here aim at studying both the two-state and the four-state protocol variants. Since the original notations of each of these are incompatible, we opt here for the standard quantum computation convention, recalled below.

The two-state protocol goes as follows (the analogous four-state protocol is presented in Section~\ref{4-state}):
\begin{enumerate}
\item Alice and Bob agree on a computational basis $\{|0\rangle , |1\rangle \}$, and, as usual, on the vectors $|+\rangle = 1/\sqrt{2} (|0\rangle + |1\rangle )$ and $|-\rangle = 1/\sqrt{2} (|0\rangle - |1\rangle )$.
\item Bob randomly chooses one of the two states $|0\rangle$ or $|+\rangle$, then prepares and sends a particle in that state to Alice. He repeats this step a number $M$ of times of his choice.
\item Alice either measures $\hat{C}_0$, for committing to $0$, on all the particles she receives, or she measures $\hat{C}_1$, for committing to $1$, on all the particles she receives, where $\hat{C}_0$ and $\hat{C}_1$ are the observables defined by
\begin{align}
\begin{split}
\hat{C}_0 &= 0\cdot |0\rangle \langle 0| + 1\cdot |1 \rangle \langle 1| \\
\hat{C}_1 &= 0\cdot |+\rangle \langle +| +1\cdot |- \rangle \langle  -|.
\end{split}
\end{align}
\item Whenever Alice wants to reveal her commitment, she sends her commitment choice and the results of her measurements back to Bob.
\item Bob either accepts or discards Alice's commitment, based on some statistical criterion of his choosing.
\end{enumerate}

Since the states sent by Bob are non-orthogonal, the outcomes of Alice's measurements follow different probability distributions, according to whether $\hat{C}_0$ or $\hat{C}_1$ was the chosen observable. Indeed, if we denote by $p_{i} (j|k)$ the probability that Alice measures the value $j$ with the observable $\hat{C}_i$ whenever Bob sends a particle in state $|k\rangle$, we have
\begin{align}
\begin{split}
\label{stats_C0}
p_0 (0|0) &= 1 \\
p_0 (1|0) &= 0 \\
p_0 (0|+) &= \frac{1}{2} \\
p_0 (1|+) &= \frac{1}{2} ,
\end{split}
\end{align}
and, similarly,
\begin{align}
\begin{split}
\label{stats_C1}
p_1 (0|0) &= \frac{1}{2} \\
p_1 (1|0) &= \frac{1}{2} \\
p_1 (0|+) &= 0 \\
p_1 (1|+) &= 1 .
\end{split}
\end{align}

Choosing either observable $\hat{C}_0$ or observable $\hat{C}_1$ thus has a clear impact on Alice's set of measurement outcomes or, in other words, her sets of statistical data, her statistical signature. Consequently, once Alice reveals her results, Bob can know whether she used observable $\hat{C}_0$ or observable $\hat{C}_1$ (see below). Evidently, measuring any other observable, or different observables on each photon, will destroy the statistical signature. By exactly how much, and whether or not Alice can take any kind of advantage out of it, is the topic of the next section.

Unfortunately, things are not as simple in any real-world implementation of the protocol, where any emission, transmission and measurement processes are inevitably subject to a certain amount of white noise. To account for such imperfections during the course of the protocol, we model the white noise as a depolarizing channel, or $r$-noise channel, $\mathcal{E}_d$, such that with probability $r$ the state becomes totally mixed, while with probability $(1-r)$ it remains unharmed. It can thus be represented as a super-operator acting on a state $\rho$ by
\begin{equation}
\mathcal{E}_d (\rho )=(1-r)\rho +r\frac{I}{2}.
\end{equation}

As expected, such a channel affects the sets of probabilities (\ref{stats_C0}) and (\ref{stats_C1}), which now become functions of $r$. We abuse notation and refer to the new set of probabilities by the same symbols, as the former will not be of use henceforth. We have
\begin{align}
\begin{split}
\label{stats_C0_noise}
p_0 (0|0) &= 1-\frac{r}{2} \\
p_0 (1|0) &= \frac{r}{2} \\
p_0 (0|+) &= \frac{1}{2} \\
p_0 (1|+) &= \frac{1}{2} ,
\end{split}
\end{align}
and
\begin{align}
\begin{split}
\label{stats_C1_noise}
p_1 (0|0) &= \frac{1}{2} \\
p_1 (1|0) &= \frac{1}{2} \\
p_1 (0|+) &= \frac{r}{2} \\
p_1 (1|+) &= 1-\frac{r}{2} .
\end{split}
\end{align}

Finally, we need to formulate a systematic way for Bob to accept or discard Alice's commitment. Once Alice reveals her commitment and her measurement results, Bob's goal is to determine whether her statistics follow the probability distribution defined by (\ref{stats_C0_noise}), the one defined by (\ref{stats_C1_noise}), or neither of these two. This problem, of deciding whether or not a sample comes from a given distribution, is an extremely well-known problem in statistics, and can be solved through a variety of methods, jointly called goodness-of-fit tests, the most famous of which is likely to be the chi-squared test. We  here, however, take advantage of the simplicity of our particular case and instead use a far more straightforward test, called the binomial test.

To do so, let us assume that Bob sends $N_0$ particles in state $|0\rangle$. Without loss of generality, suppose Alice measures the $\hat{C}_0$ observable on all of them, i.e., she commits to the value $0$. Then the probability $P(n_0 \mbox{ } 0\text{'s})$ that she obtains $n_0$ times the value $0$ after measuring all the particles is given, by definition, by a binomial distribution
\begin{equation}
P(n_0 \mbox{ } 0\text{'s}) = \binom{N_0}{n_0} p_0^{n_0} (0|0) p_0^{N_0-n_0} (1|0).
\end{equation}
This binomial distribution has mean value $\mu_0$ and variance $\sigma_0^2$ defined by
\begin{align}
\begin{split}
\mu_0 &= N_0 p_0 (0|0) \\
\sigma_0^2 &= N_0 p_0 (0|0) (1-p_0 (0|0)).
\end{split}
\end{align}
The binomial test then goes as follows: Bob proceeds if Alice's statistics satisfy
\begin{equation}
\label{cond_0}
n_0 \in [\mu_0 -3\sigma_0 , \mu_0 + 3\sigma_0 ] .
\end{equation}
Otherwise, he immediately aborts the protocol. The above condition ensures that if Alice is committing to $0$, then she has $\sim \!99.7\%$ chance of having her first set of statistics accepted by Bob. This value is simply
\begin{equation}
0.997 \sim \sum_{k=\lceil \mu_0 -3\sigma_0 \rceil}^{\lfloor \mu_0 +3\sigma_0 \rfloor} \binom{N_0}{k} p_0^{k} (0|0) p_0^{N_0-k} (1|0).
\end{equation}
Analogously, Bob tests Alice's statistics whenever state $|+\rangle$ is sent, and proceeds if and only if
\begin{equation}
\label{cond_1}
n_+ \in [\mu_+ -3\sigma_+ ,\mu_+ +3\sigma_+ ] .
\end{equation}
Otherwise, he immediately aborts. The probability that Alice's commitment is accepted by Bob, i.e. the probability that Alice passes both tests, is then simply the product of the separate probabilities. Note, of course, that one could choose a standard deviation factor other than $3$. The smaller it is, the more restrictive the criterion becomes. For simplicity, throughout the paper we set that Bob sends an equal number of particles in each state, i.e., $N_0 = N_+ = N$ (and $N_{0/1} = N_{+/-}$ in the four-state case), the assumption secured by the law of large numbers.

\begin{figure}[H]
\centering
\includegraphics[width=0.6\textwidth]{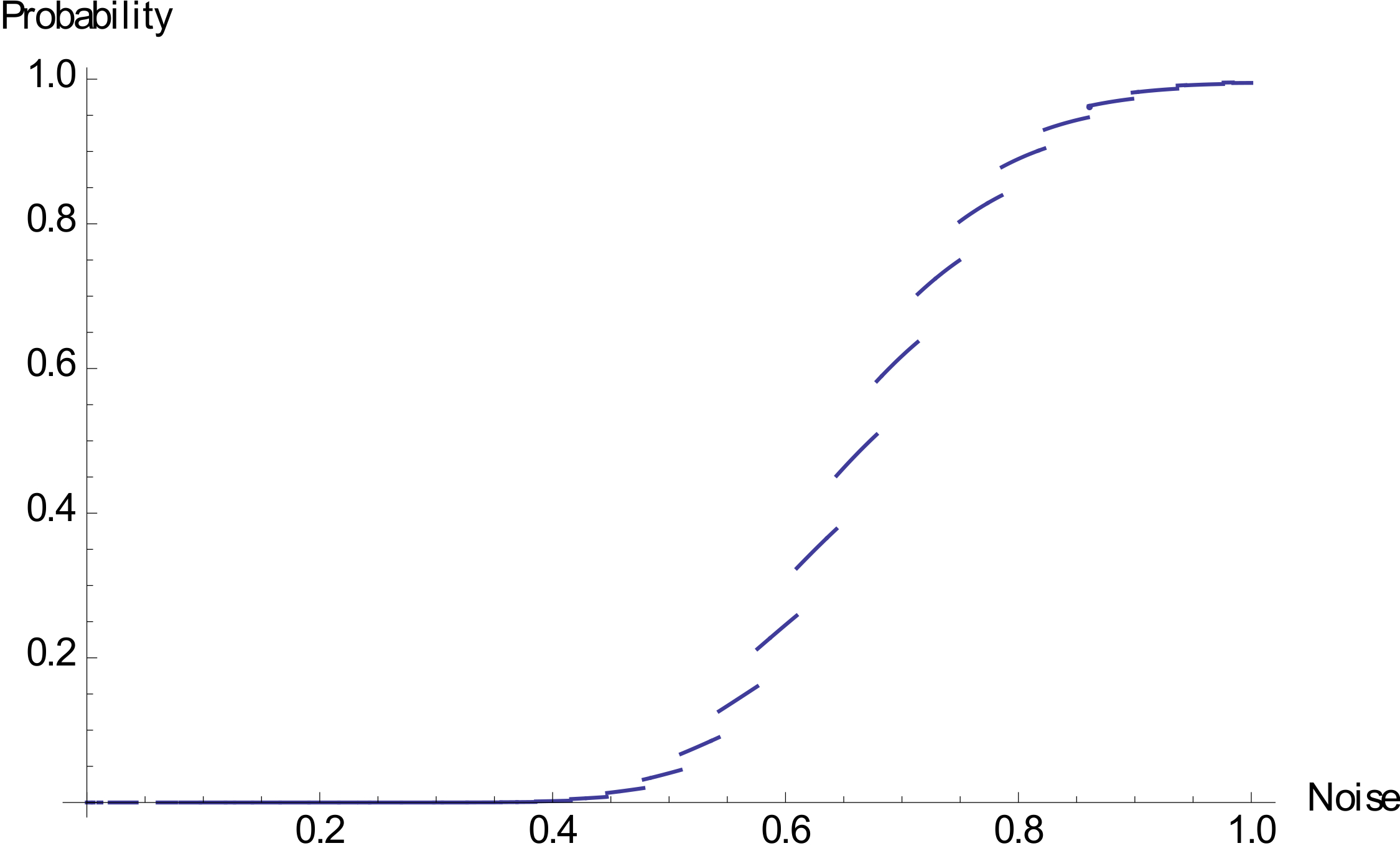}
\caption{Probability, as a function of white noise, of an honest Alice committing to $1$, with $M = 100$ successful measurements, passing Bob's test of commitment to $0$.}
\label{fig1}
\end{figure}

The test is, \textit{mutatis mutandis}, the same if Alice were to commit to $1$ instead (Fig.~\ref{fig1}). However, there is still a loose end: there is a nonzero probability that an Alice trying to commit to $1$ produces statistics that satisfy both \eqref{cond_0} and \eqref{cond_1}, thus convincing Bob that she committed to $0$, and making the test nonviable. This probability decreases exponentially with the number of measurements. Heuristically, a number of total measurements of $M = 100$ (about fifty 0's measured, and fifty 1's measured, $N = N_0 = N_+ = 50$) is enough to guarantee viability of the protocol for any reasonable amount of noise, as Fig.~\ref{fig1} shows.

\section{Optimal cheating observable}
\label{opt_cheat}

The main goal of this section is to answer a simple question: Can Alice cheat? In other words, can she devise a strategy so that she is able to convince Bob that she made a choice when in reality she did not? Naturally, Bob could also try to break the protocol. However, he has a much tougher job. For Bob to break the protocol, he would need to know what Alice's choice of measurement was before Alice makes it public. Since he has no access to Alice's laboratory, doing so is impossible, even if using entangled pairs, due to nonsignaling. Alice's side of things seem much more promising. For Alice to cheat, she has to be able to produce statistics that follow either distribution \eqref{stats_C0_noise} or distribution \eqref{stats_C1_noise} upon request. Since the only thing she can do is perform measurements on the incoming particles, she needs to find the measurement that best allows her to mimic \emph{both} those distributions, eventually after some post-processing.

Note, of course, that, in an ideal setting, Alice could simply store the incoming particles in a quantum memory and measure them only when announcing her commitment. However, stable long-term quantum memories are still a long way from becoming an everyday item, and we thus assume that Alice is forced to perform her measurements on the incoming particles as soon as they arrive. For a more complete discussion, including the case of noisy memories, please refer to \cite{Loura14}.

\subsection{The two-state protocol}

The first attempt at creating a cheating strategy is, naturally, to try and guess the state of every photon sent by Bob. Since there is no way of perfectly and systematically distinguishing non-orthogonal states, the best Alice can do, provided she is constrained to performing single-qubit measurements only, is to measure an observable ``in-between'' $\hat{C}_0$ and $\hat{C}_1$. Finding the basis that best discriminates states $|0\rangle$ and $|+\rangle$ is a common and well-studied problem in quantum mechanics. Such a basis is called a Breidbart basis. Specifically, let $|\tilde{0} \rangle$ ($|\tilde{1} \rangle$) denote the vector $|0 \rangle$ ($|+ \rangle$) rotated by $-\pi /8$ ($\pi/8)$, so as to form the aforementioned Breidbart basis, and define a new observable, the {\em cheating observable} $\hat{C}_{ch}$, by
\begin{equation}
\hat{C}_{ch} = 0\cdot |\tilde{0} \rangle \langle \tilde{0} | + 1\cdot |\tilde{1} \rangle \langle \tilde{1} |,
\end{equation}
where
\begin{align}
\begin{split}
|\tilde{0} \rangle &= \cos \frac{\pi}{8} |0 \rangle -\sin \frac{\pi}{8} |1 \rangle \\
|\tilde{1} \rangle &= \sin \frac{\pi}{8} |0\rangle + \cos \frac{\pi}{8} |1 \rangle .
\end{split}
\end{align}
Armed with this observable, Alice has a certain probability $\mathfrak{p}_{ch} (a|b)$ of obtaining the result $a$ whenever state $|b \rangle$ is sent. Concretely, in an $r$-noise channel, we have
\begin{align}
\begin{split}
\mathfrak{p}_{ch} (0|0) &= \frac{r}{2} + \frac{1}{4} (2+\sqrt{2}-(2+\sqrt{2})r) \\
\mathfrak{p}_{ch} (1|0) &= \frac{r}{2} +\frac{1}{4} (2-\sqrt{2} -(2-\sqrt{2})r) \\
\mathfrak{p}_{ch} (0|+) &= \frac{1}{4} (2-\sqrt{2}+r\sqrt{2}) \\
\mathfrak{p}_{ch} (1|+) &= \frac{1}{4} (2+\sqrt{2} -r\sqrt{2} ). 
\end{split}
\end{align}
Note that the Breidbart basis, optimal for distinguishing states $|0\rangle$ and $|+\rangle$, is also optimal for distinguishing the mixed states obtained as a result of the action of a depolarizing channel (due to the symmetry of white noise) as it saturates the Helstrom bound.

Once Alice has inferred the state of every photon, she can send her results back to Bob in hopes of passing his test. Such a strategy has been considered and analysed in previous work \cite{Loura14}. However, she can do better. After all, Alice knows that her inference is not completely accurate, and so sending her inference on what the states sent by Bob may be is definitely not the best she can do. In fact, even if Alice were able to perfectly determine the state of Bob's particles, she would not send these results back to Bob but, rather, results that would emulate the statistics of an honest Alice [such as \eqref{stats_C0} or \eqref{stats_C0_noise}]. She can thus proceed as follows: every time she obtains a $0$ with the cheating observable, she decides either to change the result, with a certain probability $p_{0\to 1}$, or to keep it, with probability $(1-p_{0\to 1})$. Analogously, every time Alice obtains a $1$, she decides either to change it, with probability $p_{1\to 0}$, or to keep it, with probability $(1-p_{1\to 0})$. After such a procedure, denoting by $p_{ch} (a|b)$ the probability that Alice sends the value $a$ back to Bob whenever he sends $|b\rangle$, we have:
\begin{align}
\begin{split}
p_{ch} (0|0) &= \mathfrak{p}_{ch} (0|0) \cdot (1-p_{0\to 1}) + \mathfrak{p}_{ch} (1|0) \cdot p_{1\to 0} \\
p_{ch} (1|0) &= \mathfrak{p}_{ch} (1|0) \cdot (1-p_{1\to 0}) + \mathfrak{p}_{ch} (0|0) \cdot p_{0\to 1} \\
p_{ch} (0|+) &= \mathfrak{p}_{ch} (0|1) \cdot (1-p_{0\to 1}) + \mathfrak{p}_{ch} (1|0) \cdot p_{1\to 0} \\
p_{ch} (1|+) &= \mathfrak{p}_{ch} (1|1) \cdot (1-p_{1\to 0}) + \mathfrak{p}_{ch} (0|1) \cdot p_{0\to 1}.
\end{split}
\end{align}

The next step is rather straightforward: determine the values of $p_{0\to 1}$ and $p_{1\to 0}$ that most increase Alice's chances of successfully cheating. Naturally, this depends on Bob's criterion, and on whether Alice is trying to trick Bob into believing she committed to $0$, or trying to trick him into believing she committed to $1$. Without loss of generality, assume that Alice is trying to forge a commitment to $0$, and that Bob is using the goodness-of-fit test described in Section \ref{prtcl_def}, that is, he accepts Alice's commitment to $0$ if the number of $0$'s revealed by Alice falls within the interval $[\mu_0 -3\sigma_0 ,\mu_0 +3\sigma_0 ]$ whenever $|0\rangle$ was sent [see \eqref{cond_0}], and the number of $1$'s within the interval $[\mu_+ -3\sigma_+ , \mu_+ +3\sigma_+]$ whenever $|+\rangle$ was sent [see \eqref{cond_1}]. The probability that a cheating Alice passes this test is then given by
\begin{equation}
\label{prob_cheat_succ}
p_{ch} = \left( \sum_{k=\mu_0 -3\sigma_0}^{\mu_0+3\sigma_0} \binom{N_0}{k} p_{ch}^k (0|0) \cdot p_{ch}^{N_0-k} (1|0) \right) \left( \sum_{k=\mu_+ -3\sigma_+}^{\mu_+ + 3\sigma_+}  \binom{N_+}{k} p_{ch}^k (0|+) \cdot p_{ch}^{N_1-k} (1|+) \right)
\end{equation}
where $N_0 = N_+ = N$ are the number of $|0\rangle$ states and the number of $|+\rangle$ states sent by Bob, respectively.

\begin{figure}[H]
	\centering
			\includegraphics[width=\textwidth]{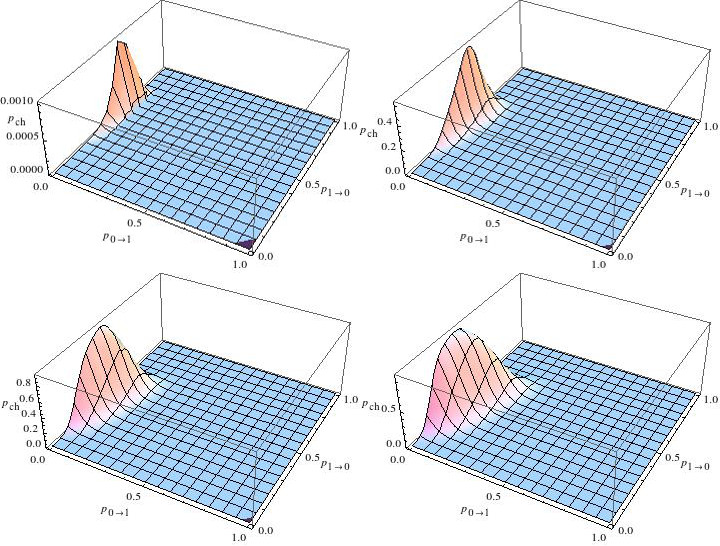}
	\caption{Probability $p_{ch}$ of a cheating Alice with $M = N_0 + N_+ = 2N = 100$ successful measurements passing Bob's test of commitment to $0$, for different values of the noise parameter $r$ ($0$ - top left, $0.08$ - top right, $0.16$ - bottom left, $0.24$ - bottom right).}
	\label{viabcheat}
\end{figure}

Although finding an expression for the maximum of such a function is non-trivial, it is quite simple to do so numerically. In Fig.~\ref{viabcheat} we present a few plots of $p_{ch}$
, for different values of the noise parameter $r$, and for $M = N_0 + N_+ = 2N = 100$.

\begin{figure}[H]
	\centering
	\includegraphics[width=0.45\textwidth]{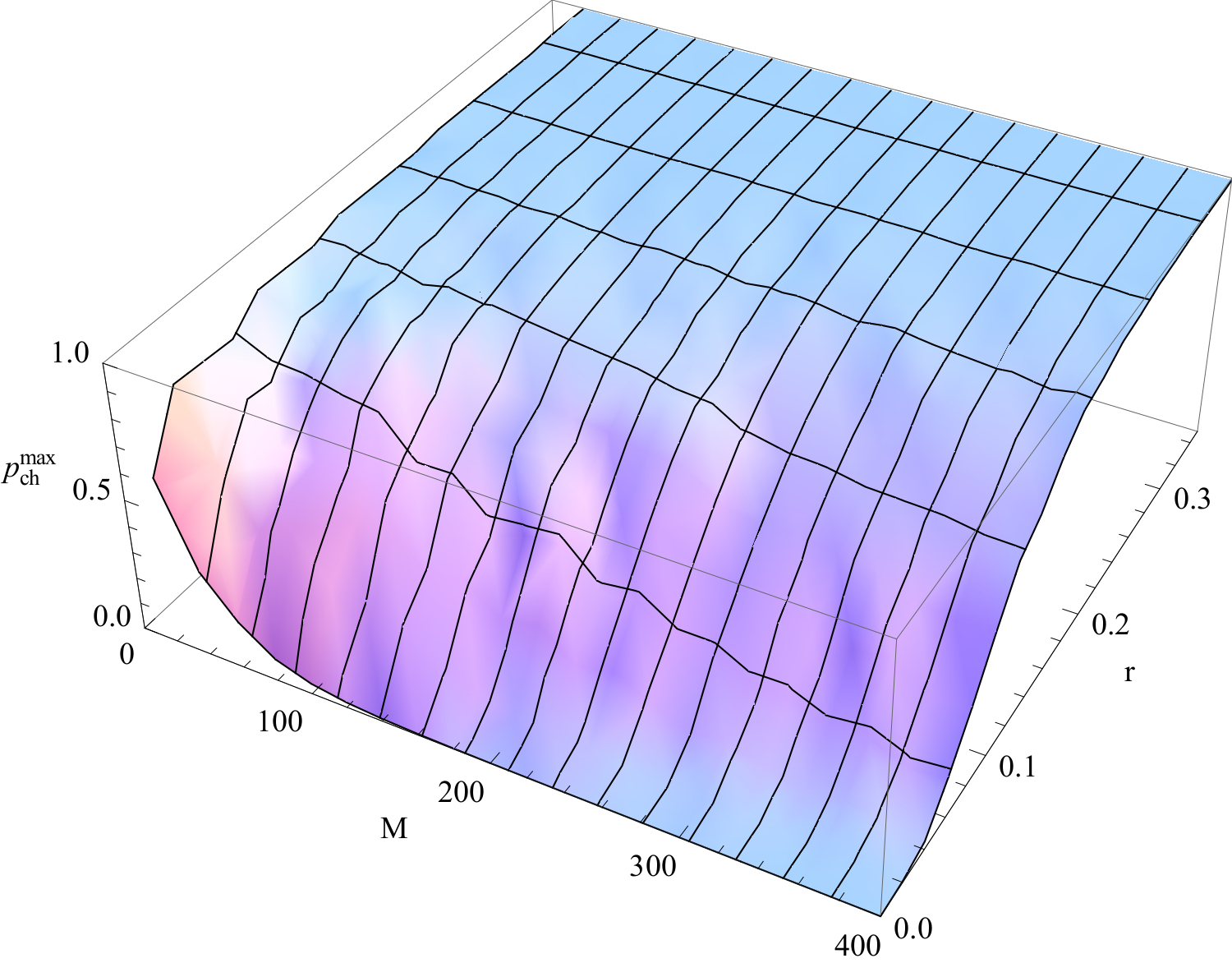}
	\caption{Probability $p_{ch}^{\text{max}}$ of a cheating Alice as a function of the number $M= N_0 + N_+ = 2N$ of successful measurements passing Bob's test of commitment to $0$, and the noise parameter $r$.}
	\label{viabcheatmax}
\end{figure}

\looseness1 As can be seen, to obtain an optimal cheating strategy in the case of a noiseless channel, a cheating Alice should always choose $p_{0\to 1}$ to be $0$. For higher values of noise Alice should choose $p_{0\to 1}$ to be slightly higher than $0$. For $M = 100$ the threshold value for $r$, for which Alice should choose $p_{0\to 1} > 0$, is above $0.52$, and it decreases as $M$ grows. As for $p_{1\to 0}$, the optimal value lies in the interval $(0.4,0.5)$, but depends on the noise parameter $r$. Given the optimal choice of the parameters  $p_{0\to 1}$ and $p_{1\to 0}$, the maximal probability of cheating, $p_{ch}^{\text{max}}$, is presented in Fig.~\ref{viabcheatmax}, as a function of $M = N_0 + N_+ = 2N$ and $r$.

\subsection{The four-state protocol}
\label{4-state}

It is quite interesting to compare the two-state cheating strategy above with the analogous cheating strategy for the four-state variant of the protocol. This variant, initially proposed in \cite{Danan12}, goes as follows:
\begin{enumerate}
\item Alice and Bob agree on the two observables
\begin{align}
\begin{split}
\hat{C}_0 &= 0\cdot |0\rangle \langle 0| + 1\cdot |1\rangle \langle 1| \\
\hat{C}_1 &= 0\cdot |-\rangle \langle -| +1\cdot |+\rangle \langle +|.
\end{split}
\end{align}
\item Bob randomly chooses one of the four states, $|0\rangle$, $|1\rangle$, $|+\rangle$ of $|-\rangle$, and then prepares and sends a particle in that state to Alice. He repeats this step a number $M$ of times of his choice.
\item Alice either measures $\hat{C}_0$, for committing to $0$, on all the particles she receives; or measures $\hat{C}_1$, for committing to $1$, on all the particles she receives.
\item Whenever Alice wants to reveal her commitment, she sends her commitment choice along with the results of her measurements back to Bob.
\item Bob either accepts or discards Alice's commitment, based on some statistical criterion of his choosing.
\end{enumerate}

\begin{figure}[H]
	\centering
	\includegraphics[width=\textwidth]{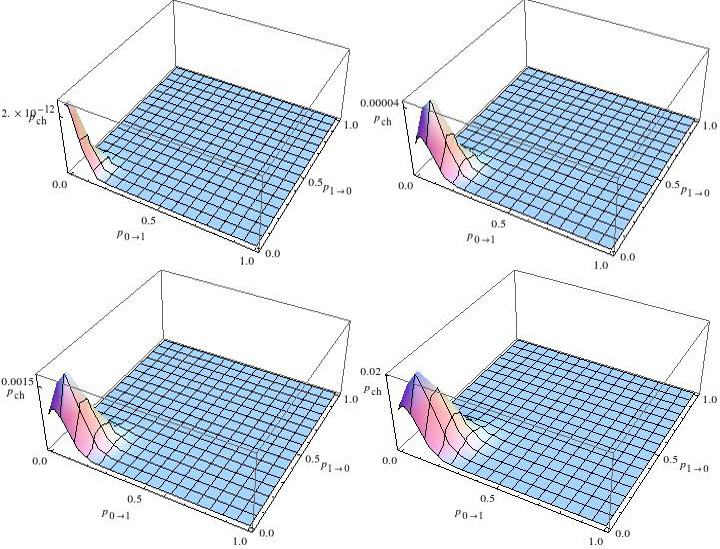}
	\caption{Probability of a cheating Alice with $M = N_0 + N_1 + N_+ + N_- = 200$ successful measurements passing Bob's test of commitment to $0$, for different values of the noise parameter $r$ ($0$ - top left, $0.08$ - top right, $0.16$ - bottom left, $0.24$ - bottom right).}
	\label{viabcheatfour}
\end{figure}

Instead of two tests, Alice now has to pass four distinct tests: one for whenever each different state was sent. The four probability distributions are now given by Equations \eqref{stats_C0_noise}, \eqref{stats_C1_noise}, and two additional but analogous ones for whenever states $|1\rangle$ and $|-\rangle$ are sent. Again, in Fig.~\ref{viabcheatfour} we plot the probability that a cheating Alice, measuring all photons in the Breidbart basis (the same as for the two-state protocol), passes all four tests of commitment to 0, as a function of the probability of changing her mind about $0$'s ($p_{0\to 1}$) and about $1$'s ($p_{1\to 0}$), for $N=N_{0/1}=N_{+/-}=50$.

\begin{figure}[H]
	\centering
	\includegraphics[width=0.45\textwidth]{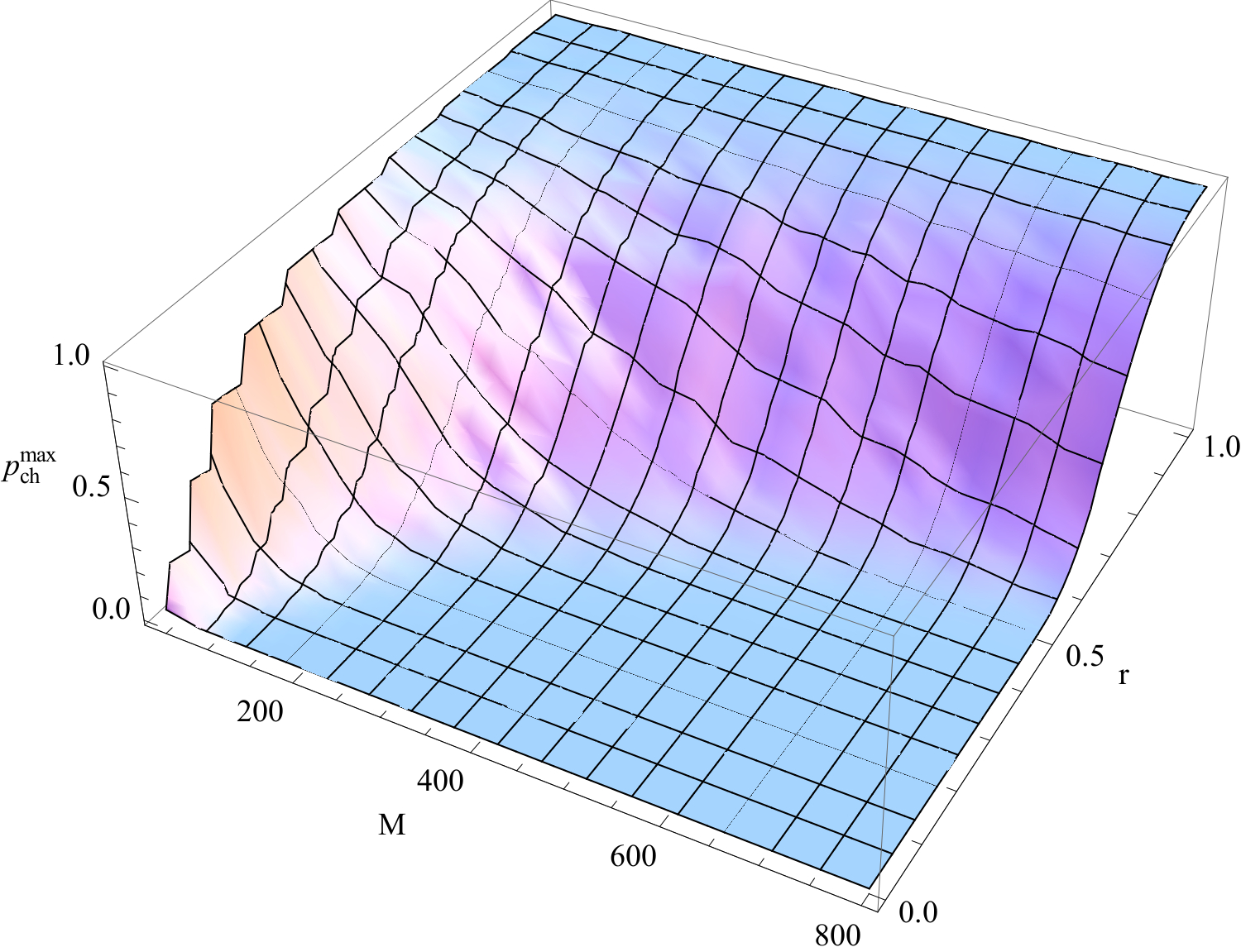}
	\caption{Probability $p_{ch}^{\text{max}}$ of a cheating Alice as a function of the number $M = 4N$ of successful measurements passing Bob's test of commitment to $0$, and the noise parameter $r$.}
	\label{viabcheatfourmax}
\end{figure}

As can be seen, it is always advantageous for a cheating Alice to change her mind about both $0$'s and $1$'s. Finally, in Fig.~\ref{viabcheatfourmax} we show the maximal probability of cheating, $p_{ch}^{\text{max}}$, as a function of $M = N_{0} + N_{1} + N_{+} + N_{-} = 4N$ and $r$. Compared with the two-state case, it is clear that in terms of the resources (the total number of measurement outcomes, $M = 2N$ for two-state, and $M = 4N^\prime$ for four-state protocols, with $N^\prime=N/2$) the four-state protocol is advantageous.

\section{Faking-distance attack}
\label{fake_dist}

In any experimental implementation of the above protocols, there will be inevitable losses during transmission, just as there is inevitable noise. These losses, just like noise, need to be taken into account by Bob when receiving the final data from Alice. Indeed, if Alice is close by, then Bob expects her to receive most of the emitted particles, say, photons, and thus expects a measurement result for almost each photon sent. On the other hand, if Alice is far away, then several photons will be lost during transmission, and thus Bob will be expecting only a small number of measurement results compared to the number of photons sent. Curiously enough, this mundane phenomenon is advantageous for a cheating Alice.

Suppose Alice is much closer to Bob than she claimed to be, meaning Bob is expecting that only a fraction $f$ of the particles sent will end up producing a measurement result. In reality, Alice is able to generate measurement results for (almost) all emitted photons. So if Alice has a way to produce statistical data in perfect agreement with those of two honest parties committing to different values for a fraction $f$ of all particles, she can trick Bob into thinking her forged statistics are the legitimate statistics obtained by an Alice far away, who had access to only $f$ times the particles sent.

Alice can achieve this very easily by simply measuring $\hat{C}_0$ on half of the received particles, and $\hat{C}_1$ on the other half. When asked to give proof of her commitment, Alice simply sends the half of the measurement results that best suits her, thus emulating an honest Alice who would have had access to only $50\%$ of the particles sent. This strategy is stronger than the equally common technique of unambiguous state discrimination, where Alice can learn the true state of around $29\%$ of the particles she receives.


A cheating Alice, making her measurements right outside Bob's laboratory, while pretending to be far away, can thus cheat if Bob is expecting that only $50\%$ or less of the particles sent end up producing a measurement result (by an honest Alice). There is thus a maximal distance up to which the protocol is secure: the distance beyond which an honest Alice is able to measure only $50\%$ or less of the particles sent. To estimate this distance, let $N_P$ be the total number of pulses produced by Bob. In each pulse, a photon is emitted following a certain probability distribution with mean $\mu$. The total number of emitted photons is then $N_E = \mu N_P$. The number of photons received by an honest Alice is thus $N_R = 10^{-\alpha L/10}N_E$, where $\alpha$ is the attenuation coefficient and $L$ is the length of the optical fibre. Finally, due to a finite detector efficiency $\eta$, the number of measurement outcomes of an honest Alice is $N_M = \eta N_R = \eta\mu 10^{-\alpha L/10}N_P$, while for a cheating Alice it is $N_{cM} = \eta N_E = \eta\mu N_P$. A cheating Alice, performing measurements right outside Bob's laboratory ($L = 0$), 
produces statistics that are compatible with her future choice of commitment on 50$\%$ of the measurement outcomes.
Note that we are assuming that a cheating Alice has the same type of detector as an honest Alice, since we are dealing with a practical case scenario. 
Thus, the maximal distance $L_{\text{max}}$ for which the protocol is still secure is given by $N_M = N_{cM}/2$, leading to
\begin{equation}
\label{max_safe_dist}
L_{\text{max}} =\frac{10}{\alpha} \log_{10}(2).
\end{equation}

For a typical value of $\alpha =0.2$ (as used in, for instance, \cite{alm:etal:15}), this amounts to a maximal safe distance of 15km.

This attack shares quite some similarity to the famous photon-number splitting attack on the BB84 protocol, which can be patched using decoy states, as suggested by Hwang in \cite{Hwang03}. Unfortunately, this remedy is not applicable in our case, since the evildoer (Eve) and the honest agent (Alice) are one and the same. We are thus left with only a maximum safe distance.

In the case of real-life noisy scenario, it is natural to assume that the ``distant'' Alice obtains imperfect results due to the ``distant'' noise factor $r_d$, while the cheating ``nearby'' Alice has noise given by $r_n < r_d$. A cheating Alice therefore has better statistics than an honest one, and can use her ``good'' results on 50$\%$ of the cases to infer, say, $(50 + \delta)\%$ of the ``less quality'' results obtained by an honest Alice: by randomly choosing $\delta \%$ of the ``missing'' results, a cheating Alice spoils the overall quality of her statistics, to match those of an honest Alice, which additionally decreases $L_{\text{max}}$. In other words, we have 

\begin{equation}
\label{N_noisy}
N_M = \frac{N_{cM}}{2} + \Delta N,
\end{equation}
where $\Delta N = \delta \cdot N$ (consequently, $\delta = 1/2 - 10^{-\alpha L/10}$). The statistics $p_{c;r_d}(\ast |\ast)$, with $c \in \{ 0,1 \}$, of an honest Alice are given by Equations~\eqref{stats_C0_noise} and~\eqref{stats_C1_noise}, with $r=r_d$. The statistics of a cheating Alice are given by a weighted sum of the ``good'' results, measured on $N_{cM}/2$ photons, and random guesses, presented for $\Delta N$ photons:
\begin{equation}
p^{ch}_{c;r_n}(\ast |\ast) = \frac{\frac{N_{cM}}{2}}{\frac{N_{cM}}{2} + \Delta N} p_{c;r_n}(\ast |\ast) + \frac{\Delta N}{\frac{N_{cM}}{2} + \Delta N}\frac{1}{2}.
\end{equation}
Equating, say $p^{ch}_{0;r_n}(0 |0)$ with $p_{0;r_d}(0 |0)$, and using~\eqref{N_noisy}, one obtains the expression for the maximum safe distance:
\begin{equation}
\label{max_safe_dist_noise}
L_{\text{max}} =\frac{10}{\alpha} [ \log_{10}(2) + \log_{10}(1 - \frac{r_d - r_n}{1-r_d})].
\end{equation}
Note that the expression~\eqref{max_safe_dist_noise} is defined only for $(r_d - r_n)/(1 - r_d) < 1$. If $ r_d \geq (r_n + 1)/2$, there exist no $L_{\text{max}}$, i.e., a dishonest Alice can always fake her results successfully, which comes as no surprise: for such high noise, $r_d > 1/2$, the two commitments become indistinguishable and the protocol is not viable. The above analysis is applicable to both two- and four-state protocols.

\section{Multi-photon sources}
\label{sing_phots}

There is still one detail that we have discarded up until now but that plagues any implementation that may currently be done: the nonexistence of perfect single-photon sources. Indeed, up until this point we have never considered the possibility that Alice may find with certainty in which state certain particles are sent by Bob. It turns out, however, that any device used to prepare photons in a specific state has a non-negligible probability of creating a pair (or even triple, quadruple, etc.) of photons in the same state instead. In a realistic scenario, one must consider a photon source whose number of photons emitted per pulse follows a Poisson distribution with a certain parameter $\mu$ (average number of photons per pulse), characteristic of the source. If we denote by $P(n,\mu )$ the probability of obtaining $n$ photons in a given pulse, we have
\begin{equation}
P(n,\mu )=\frac{\mu^n}{n!} e^{-\mu} .
\end{equation}

Alas, as soon as Bob sends more than one photon in a single pulse, which happens with probability $(1-e^{-\mu} -\mu e^{-\mu})$, then Alice may split up the photons and measure the two observables, $\hat{C_0}$ and $\hat{C}_1$, on two of them separately, thus learning the photon's true state (provided Bob follows the protocol) and producing statistics compatible with either a commitment to $0$ or a commitment to $1$ upon request. If Alice receives only one photon, she reverts to the strategy described in Section \ref{opt_cheat}. We must then once again reevaluate the sets of cheating probabilities in order to obtain $p_{CH} (i|j)$ - the probability that  a cheating Alice, using every single tool at her disposal, will send the value $i$ back to Bob, every time he sends state $|j\rangle$. When trying to forge a commitment to 0, the corresponding set of conditional probabilities for a two-state case is (the additional conditional probabilities for a four-state case are obtained analogously)
\begin{align}
\begin{split}
\label{stats_C0_cheat_final}
p_{CH} (0|0) &= \frac{1}{1-e^{-\mu}} [ \mu e^{-\mu} p_{ch}(0|0) + (1-e^{-\mu} - \mu e^{-\mu})  p_0 (0|0) ] \\
p_{CH} (1|0) &= \frac{1}{1-e^{-\mu}} [ \mu e^{-\mu} p_{ch} (1|0) + (1-e^{-\mu} - \mu e^{-\mu})  p_0 (1|0) ] \\
p_{CH} (0|+) &= \frac{1}{1-e^{-\mu}} [ \mu e^{-\mu} p_{ch} (0|+) + (1-e^{-\mu} - \mu e^{-\mu}) p_0(0|+)] \\
p_{CH} (1|+) &= \frac{1}{1-e^{-\mu}} [ \mu e^{-\mu} p_{ch} (1|+) + (1-e^{-\mu} - \mu e^{-\mu}) p_0 (1|+)],
\end{split}
\end{align}
and analogously for the case of a commitment to 1, by replacing $p_0 (\ast |\ast)$ with $p_1 (\ast |\ast)$.

In more detail, take for instance the very first value: $p_{CH} (0|0)$. This value has meaning only if Bob actually sent something, which happens with probability $(1-e^{-\mu})$, hence the factor $\frac{1}{1-e^{-\mu}}$. Given the case, two mutually incompatible results may occur. Either Bob sends a single particle, which happens with probability $\mu e^{-\mu}$, and Alice measures it using her single-photon strategy from Section~\ref{opt_cheat} [hence $p_{ch} (0|0)$], or Bob sends two or more particles, which happens with probability $(1-e^{-\mu} -\mu e^{-\mu} )$, and Alice perfectly emulates the probability distribution of an honest party [hence $p_0 (0|0)$]. All other expressions follow an analogous reasoning.

Note that that the conditional probabilities $p_{ch} (\ast | \ast)$, corresponding to the strategy in Section~\ref{opt_cheat}, are those which maximise the overall cheating probability $p_{ch}^{max}$ with respect to the case of the {\em multi-photon} sources. Thus, the optimal values of $p_{0\to 1}$ and $p_{1\to 0}$ are, in general, different for the strategies that maximise the cheating probability for the case of the {\em single-photon} source, and the case of the {\em multi-photon} source. In Tables~\ref{tab:2-state_opt} and~\ref{tab:4-state_opt}, we compare the mentioned optimal values for both two- and four-state protocols, respectively, for $M = 100, 200, 300$ and $400$, with $r=0.1$ and, for the multi-photon case, $\mu = 0.2$ (the value used in the experimental realisation of a two-state protocol~\cite{alm:etal:15}).


\begin{table}
\centering
\begin{tabular}{*{8}{c}}
\hline & $M$ & & & Single-photon & & Multi-photon & \\
\hline & $100$ & & & $(0.0,\,0.489663)$ & & $(0.0,\,0.492572)$ & \\
\hline & $200$ & & & $(0.0,\,0.490563)$ & & $(0.0,\,0.494314)$ & \\
\hline & $300$ & & & $(0.0,\,0.479936)$ & & $(0.0,\,0.483053)$ & \\
\hline & $400$ & & & $(0.0,\,0.47544)$ & & $(0.0,\,0.478355)$ & \\
\hline
\end{tabular}
\caption{The two-state protocol's optimal values $(p_{0\to 1},p_{1\to 0})$ for single-photon (left column) and multi-photon (right column) sources, for $M = 100, 200, 300$ and $400$, with $r=0.1$ and, for the multi-photon case, $\mu = 0.2$.}
\label{tab:2-state_opt}
\end{table}

\begin{table}
\centering
\begin{tabular}{*{8}{c}}
\hline & $M$ & & & Single-photon & & Multi-photon & \\
\hline & $100$ & & & $(0.0658591,\,0.0658591)$ & & $(0.0470355,\,0.0470355)$ & \\
\hline & $200$ & & & $(0.0793028,\,0.0793028)$ & & $(0.0592754,\,0.0592754)$ & \\
\hline & $300$ & & & $(0.0939039,\,0.0939039)$ & & $(0.0743531,\,0.0743531)$ & \\
\hline & $400$ & & & $(0.101166,\,0.101166)$ & & $(0.0818707,\,0.0818707)$ & \\
\hline
\end{tabular}
\caption{The four-state protocol's optimal values $(p_{0\to 1},p_{1\to 0})$ for single-photon (left column) and multi-photon (right column) sources, for $M = 100, 200, 300$ and $400$, with $r=0.1$ and, for the multi-photon case, $\mu = 0.2$.}
\label{tab:4-state_opt}
\end{table}

The cheating strategy considered here is beyond the capabilities of technology, both at present and in any foreseeable future. To perform this kind of attack, a cheating Alice needs far more than just a nondemolition measurement. She needs a device capable of detecting whether or not the photon source emitted a single photon, and needs to change her measurement apparatus accordingly. If a single photon was sent, she measures it using the optimal cheating observable. Otherwise, she splits two of the incoming photons into two different measurement apparatuses (measuring $\hat{C}_0$ and $\hat{C}_1$ respectively). Such technology is far beyond our current capabilities.
%
More realistic is to perform the beam-splitter strategy for all of the pulses. This way, a cheating Alice would on average measure $\hat{C}_0$ on half of the single-photon pulses ($\frac 1 2 \mu e^{-\mu}N_P$), and $\hat{C}_1$ on the other half, while for the rest of the multiphoton pulses she will have the results of both the observables. Thus, for photons on which the wrong observable was measured, a cheating Alice has to add random results. When committing to $0$, her statistics for a two-state case (the additional conditional probabilities for a four-state case are obtained analogously) are given by
\begin{equation}
\label{BS_attack}
p_{BS}(\ast |\ast) = \left(1 - \frac{1}{2}\frac{\mu e^{-\mu}}{1-e^{-\mu}} \right) p_{0}(\ast |\ast) + \left(\frac{1}{2}\frac{\mu e^{-\mu}}{1-e^{-\mu}}\right)\cdot \frac{1}{2},
\end{equation}
and analogously for a commitment to $1$, by replacing $p_0 (\ast |\ast)$ with $p_1 (\ast |\ast)$.

\begin{figure}[H]
	\centering
	\includegraphics[width=\textwidth]{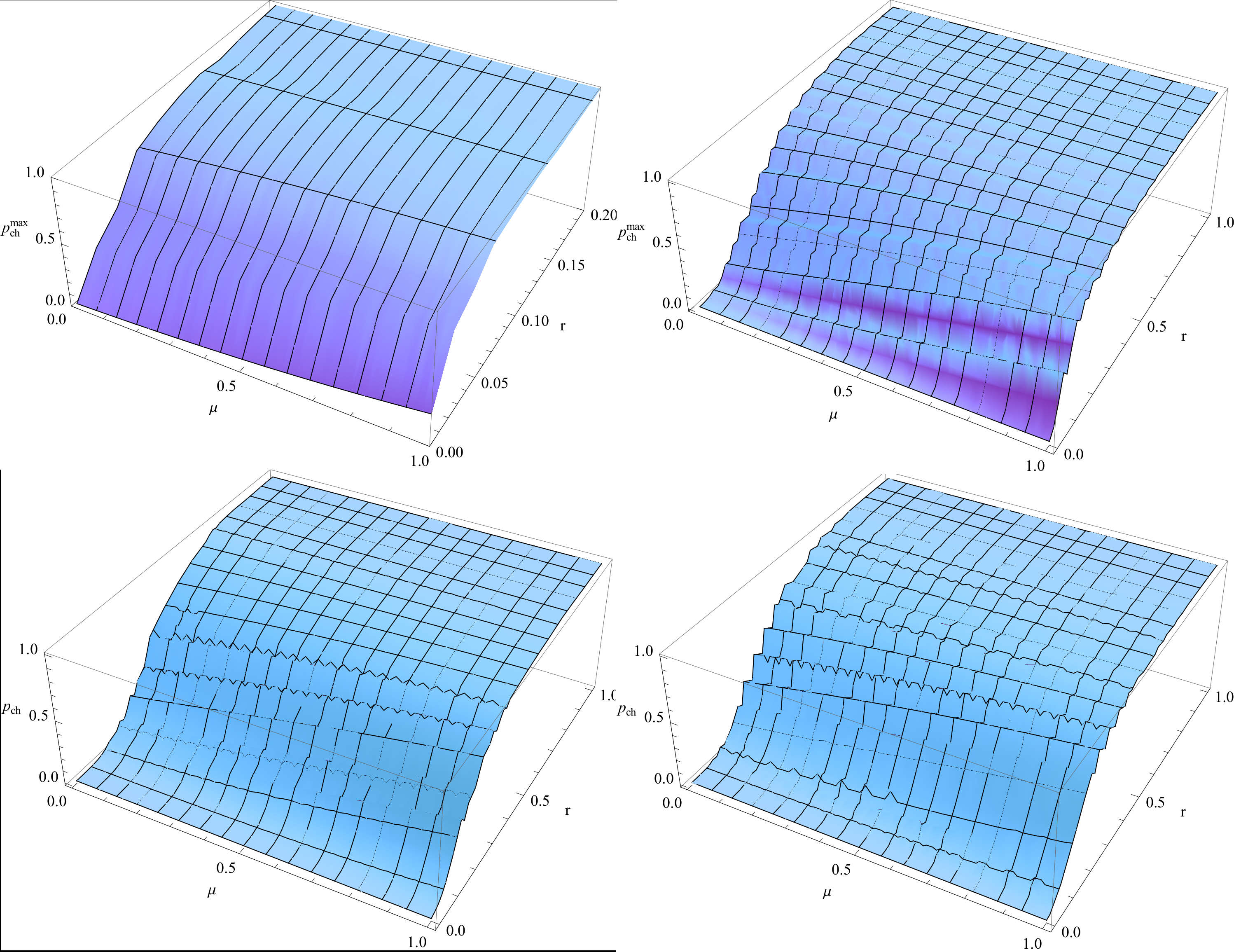}
	\caption{Probabilities of a cheating Alice with $M = 100$ successful measurements passing Bob's test of commitment to $0$, as functions of the average number of photons per pulse $\mu \in \{ 0, 1 \}$ and the noise parameter $r$, for: the two-state protocol with the less realistic strategy~\eqref{stats_C0_cheat_final} (top left); the two-state protocol with the more realistic strategy~\eqref{BS_attack} (bottom left); the four-state protocol with the less realistic strategy~\eqref{stats_C0_cheat_final} (top right); and the four-state protocol with the more realistic strategy~\eqref{BS_attack} (bottom right).}
	\label{comp_M100}
\end{figure}

\begin{figure}[H]
	\centering
	\includegraphics[width=1.30\textwidth]{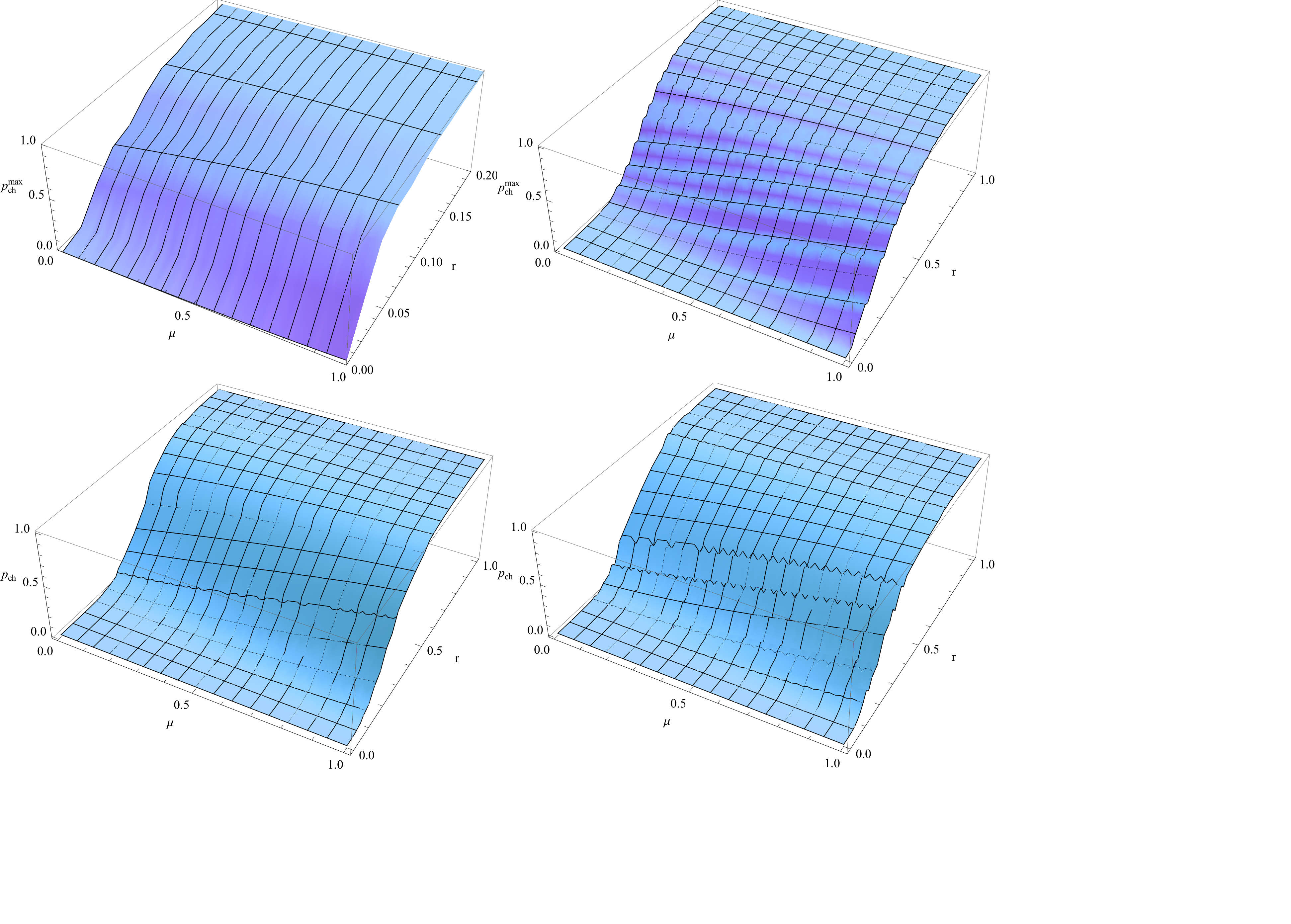}
	\caption{Probabilities of a cheating Alice with $M = 200$ successful measurements passing Bob's test of commitment to $0$, as functions of the average number of photons per pulse $\mu \in \{ 0, 1 \}$ and the noise parameter $r$, for: the two-state protocol with the less realistic strategy~\eqref{stats_C0_cheat_final} (top left); the two-state protocol with the more realistic strategy~\eqref{BS_attack} (bottom left); the four-state protocol with the less realistic strategy~\eqref{stats_C0_cheat_final} (top right); and the four-state protocol with the more realistic strategy~\eqref{BS_attack} (bottom right).}
	\label{comp_M200}
\end{figure}

\begin{figure}[H]
	\centering
	\includegraphics[width=\textwidth]{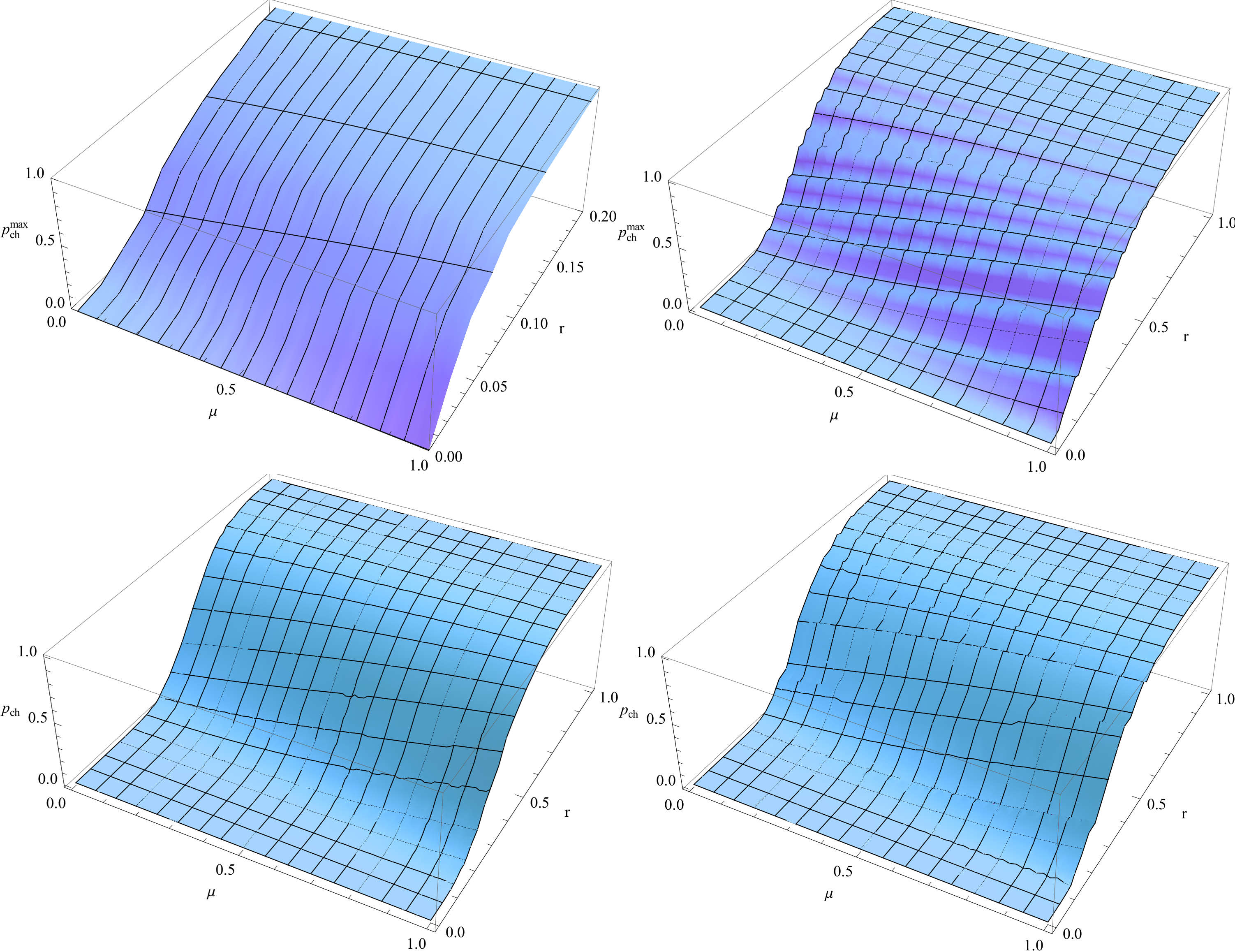}
	\caption{Probabilities of a cheating Alice with $M = 300$ successful measurements passing Bob's test of commitment to $0$, as functions of the average number of photons per pulse $\mu \in \{ 0, 1 \}$ and the noise parameter $r$, for: the two-state protocol with the less realistic strategy~\eqref{stats_C0_cheat_final} (top left); the two-state protocol with the more realistic strategy~\eqref{BS_attack} (bottom left); the four-state protocol with the less realistic strategy~\eqref{stats_C0_cheat_final} (top right); and the four-state protocol with the more realistic strategy~\eqref{BS_attack} (bottom right).}
	\label{comp_M300}
\end{figure}

\begin{figure}[H]
	\centering
	\includegraphics[width=\textwidth]{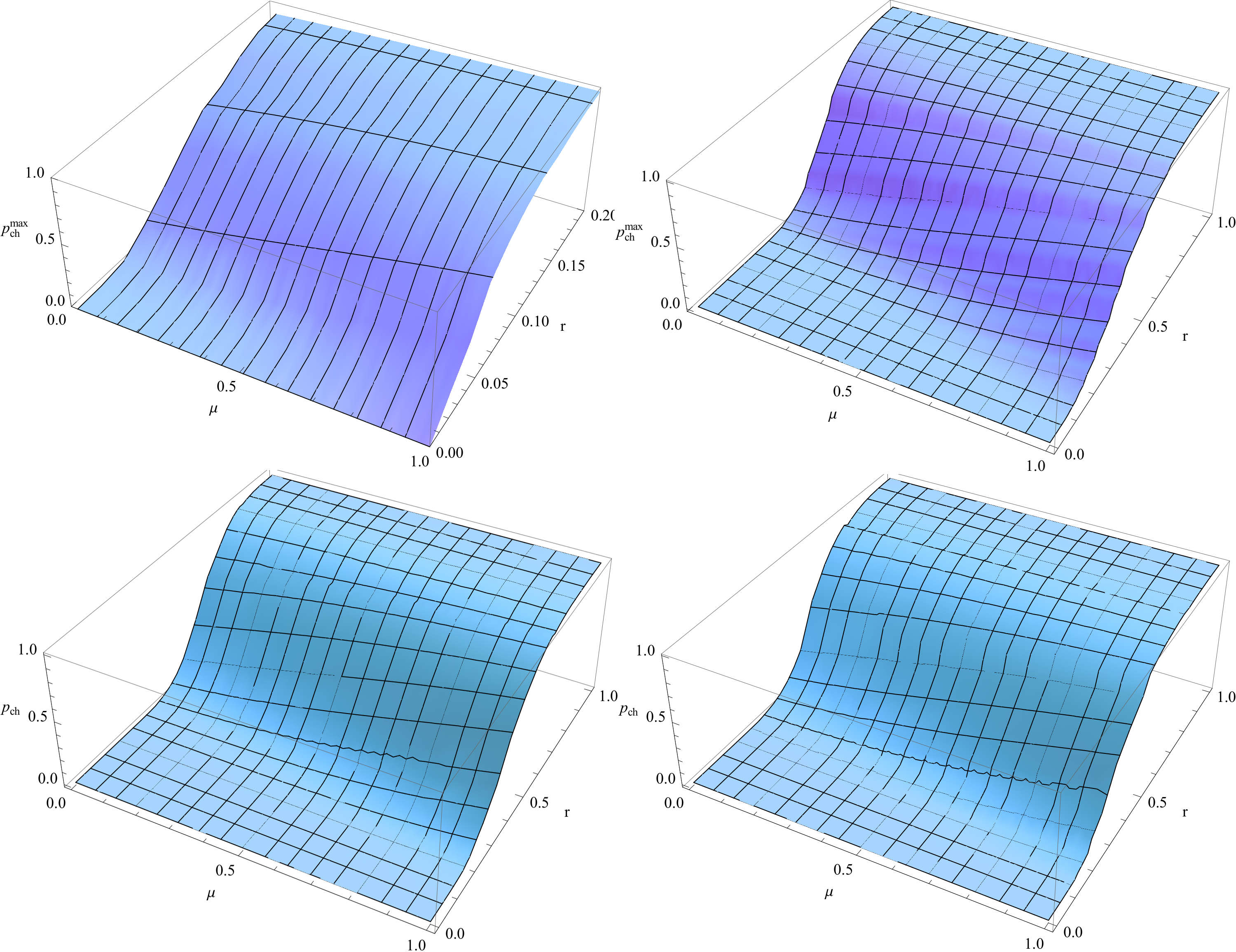}
	\caption{Probabilities of a cheating Alice with $M = 400$ successful measurements passing Bob's test of commitment to $0$, as functions of the average number of photons per pulse $\mu \in \{ 0, 1 \}$ and the noise parameter $r$, for: the two-state protocol with the less realistic strategy~\eqref{stats_C0_cheat_final} (top left); the two-state protocol with the more realistic strategy~\eqref{BS_attack} (bottom left); the four-state protocol with the less realistic strategy~\eqref{stats_C0_cheat_final} (top right); and the four-state protocol with the more realistic strategy~\eqref{BS_attack} (bottom right).}
	\label{comp_M400}
\end{figure}

Naturally, the higher $\mu$ is, the closer the two distributions, given by~\eqref{stats_C0_cheat_final} and~\eqref{BS_attack}, are to those obtained by an honest Alice committing to $0$ or committing to $1$, respectively. Thus, we consider only weak sources with a low average photon number $\mu \in \{ 0, 1 \}$ (indeed, in a recent experimental realisation of a two-state protocol~\cite{alm:etal:15}, the value $\mu \approx 0.2$ was used). In Figures~\ref{comp_M100},~\ref{comp_M200},~\ref{comp_M300} and~\ref{comp_M400} are the plots of the probabilities of cheating for both types of strategies [less realistic,~\eqref{stats_C0_cheat_final}, and more realistic,~\eqref{BS_attack}], and both versions of the protocol (two- and four-state), as functions of the average number of photons per pulse $\mu$ and the noise parameter $r$, for $M = 100, 200, 300$ and $400$, respectively.

\section{Conclusions}

We have analysed the cheating strategies for the practical two-state and four-state bit-commitment protocols presented in~\cite{Loura14} and~\cite{Danan12}, respectively. We showed that by introducing the ``post-processing'' of the raw experimental data, one can improve the na\"{i}ve cheating strategy based on the measurement in the Breidbart basis, originally studied in~\cite{Loura14}. We have studied the two cases of perfect single-photon and realistic multi-photon sources, showing that the ``post-processing'' optimal parameters $p_{0\to 1}$ and $p_{1\to 0}$ are in general different for the two cases. We have also analysed the more realistic multi-photon source beam-splitter strategy which does not rely on (single-photon) non-demolition measurements. In all of the mentioned cases, the four-state protocol shows a clear advantage over the two-state version, with respect to the resources needed to achieve the same level of security. Finally, we have analysed an alternative strategy, the {\em faking-distance attack}, based on the forged geographical location of the committing party (Alice), showing that for today's typical equipment the protocol is secure up to a distance of 15km between the parties, for both two- and four-state protocols. Our approach is, following the analysis presented in~\cite{Loura14}, straightforward to apply in the cases of noisy and/or bounded memories.

\section*{Acknowledgments}
The authors acknowledge financial support by the bilateral scientific co-operation between Portugal and Serbia through the project ``Noise and measurement errors in multi-party quantum security protocols'', No. 451-03-01765/2014-09/04 supported by the Foundation for Science and Technology (FCT), Portugal, and the Ministry of Education, Science and Technological Development of the Republic of Serbia.
R.L. thanks the FCT for support through the Ph.D. Grant No. SFRH/BD/79571/2011.
N.P. acknowledges	the support of the Security and Quantum Information Group (SQIG) and IT project QbigD funded by FCT Grants. No. PEst-OE/EEI/LA0008/2013 and No. UID/EEA/50008/2013. 
D.A., D.P. and S.P. acknowledge the support of Ministry of Education, Science and Technological Development of the Republic of Serbia, Contracts No. 171006, No. 171017, No. 171020, No. 171038 and No. 45016.

\bibliography{cheating_in_BC_protocol}{}
\bibliographystyle{plain}

\end{document}